\documentstyle[12pt]{article}
\normalbaselines
\begin{document}
\bibliographystyle{unsrt}
\vbox {\vspace{6mm}}

\begin{center}
{\large  \bf Shannon-Wehrl entropy \\ [2mm]
for cosmological and black-hole squeezing}\\[7mm]
Haret Rosu\\
{\it Instituto de F\'{\i}sica, Universidad de Guanajuato,\\
Apdo Postal E-143, 37150 Le\'on, Gto, M\'exico}\\[5mm]
Marco Reyes\\
{\it Departamento de F\'{\i}sica, CINVESTAV, Apdo Postal 14-740,\\
07000 M\'exico D.F., M\'exico}\\[5mm]
\end{center}
\vspace{2mm}

\begin{abstract}
We discuss the Shannon-Wehrl entropy within the squeezing vocabulary for
the cosmological and black hole particle production.
\end{abstract}

Models and concepts of quantum optics have been applied to quantum
cosmology (cosmological particle production) already in the seventies
\cite{lit}.
More recently Grishchuk and Sidorov [GS] \cite{gs} used the formalism of
squeezed states to discuss the spectrum of relic gravitons from inflation,
the spectrum of primordial density perturbations, and even the Hawking
radiation of Schwarzschild black holes.
Apparently there is no new physics entailed \cite{a}.
However, the squeezing language may well be more effective in characterizing
those physical processes which are of basic theoretical interest. Therefore
many authors started to use this language in the cosmological context.

Here we apply the Shannon-Wehrl entropy ($S_{sw}$) \cite{w} to the squeezing
approach of [GS] \cite{r}.

In quantum optics $S_{sw}$ is known as an important
parameter which is employed to distinguish among various types of coherent
states, measuring the relative degree of squeezing with respect to pure
coherent states for which $S_{sw}=1$ is minimum \cite{l}. It is defined
as follows
$$ S_{sw}=-\frac{1}{\pi}\int d^{2}\alpha Q(\alpha)\ln Q(\alpha) \eqno(1)$$
where $Q(\alpha)$ is the $Q$ representation of the density operator satisfying
the normalization condition $\frac{1}{\pi}\int d^{2}\alpha Q(\alpha)=1$.

The calculation of $S_{sw}$ for various types of states is not difficult
\cite{o}. Here we quote two results of which we shall make use in the
following. For the one-mode squeezed states
$$S_{sw}=1+\frac{1}{2}\ln (\sinh ^{2} r-|\epsilon |^{2}) \eqno(2)$$
where $\epsilon$ is the coherent percent component of the squeezed state.
$\epsilon=0$ means the squeezed vacuum state. For the thermal states the
$S_{sw}$ parameter may be written
$$S_{sw}=1-\ln (1-\xi)  \eqno(3)$$
where $\xi$ is the inverse of the Boltzmann modal factor
$\xi=\exp(-\beta \hbar \omega)$.

Let us pass now to the squeezing approach of [GS]. The main idea is that
gravitons created from zero point fluctuations of an initial vacuum
cosmological state are at present in an one-mode squeezed quantum state as the
result of the parametric amplification due to the interaction with the
variable gravitational background. The squeezing coefficient $r$ is a
function of the cosmological evolution. Most authors \cite{ev} use an
expansion in three stages: inflationary (i), radiation-dominated (r), and
matter-dominated (m), with the transitions between stages considered in the
`sudden' approximation in which the kinematic effects of the transitions are
neglected. Thus the Universe remains in the same quantum state as before
transitions \cite{k}, and only a redistribution (squeezing) of the
quasiparticles takes place.
The parametric amplification occurs mainly
at the inflationary stage, where the variation of the background is most
rapid. The squeezing coefficient can be obtained from the
ratios of the dimensionless scale factors at the returning (either at r-stage
or m-stage) and exit of a given mode out of the Hubble sphere at the i-stage,
as follows
$$\exp r =a(\eta _{ret})/a(\eta _{ex})  \eqno(3)$$
According to [GS] $r$ increased from $r\sim 1$ up to $r\sim 100$ for waves
with present-day frequencies ranging from $\nu \approx 10^{-8}-10^{-16}$Hz,
which were amplified at the inflationary stage only.
For waves in the range $\nu \approx 10^{-16}-10^{-18}$Hz, the squeezing
parameter may reach a value of 120 due to the additional amplification at
the matter-dominated transition. We see that cosmological squeezing is
about two orders of magnitude bigger than ordinary laboratory squeezing.
This is indicative of the huge mean number of quasiparticles in every mode.
Making use of the numerical values of the cosmological squeezing coefficient
we can plot the $S_{sw}$ entropic parameter for the one-mode squeezed graviton
states according to Eq.(2).

\vskip 6.1cm

\begin{quotation}
Fig. 1: Shannon-Wehrl entropy for graviton squeezed states with different
coherent components $\epsilon$ (full line $\epsilon$ = 0\%, slash line
$\epsilon$ = 10\%, slash-dot line $\epsilon$ = 20\%).
\end{quotation}

The non-zero coherent component we allowed for would correspond to possible
deviations of the initial quantum state of gravitons from the vacuum state
\cite{gr}.

In the case of Schwarzschild black holes a two-mode squeezing comes into
play for any type of radiation detected at asymptotic infinity. However
due to the special causal structure of the black hole spacetime, the
asymptotic observations are limited to one mode only. Under such
conditions the detected states turn into thermal ones. Actually, Hawking
radiation is a distorted blackbody radiation, but we can consider it as
an effective thermal one \cite{bek}. Therefore we plotted $S_{sw}$
according to Eq.(3), with $\gamma$ in the effective Boltzmann factor
defined by
$$\frac{1}{\exp(\gamma)-1}=
\frac{\Gamma (\omega)}{\exp(\beta _{h}\hbar\omega)-1} \eqno(4)$$
where $\beta _{h}$ is the horizon inverse temperature parameter, and
$\Gamma  (\omega)$ is the penetration factor of the curvature and angular
momentum barrier around the black hole.

\vskip 6cm

\begin{quotation}
Fig. 2: Shannon-Wehrl entropy for the `thermal' radiation of a $M=10^{17}$ g
Schwarzschild black hole as a function of the inverse of the effective
Bolzmann factor.
\end{quotation}
We chose the mass of the black hole so that no massive particles are emitted.
This would better correspond to the analogy with quantum optics.

\section*{Acknowledgments}
This work was supported partially by CONACyT Grant No. F246-E9207 to the
University of Guanajuato, and by a CONACyT Graduate Fellowship.

\end{document}